\documentclass[letter]{aa} 

\usepackage{hyperref}
\usepackage{pdflscape}
\usepackage{xcolor}
\usepackage{lscape}

\usepackage{graphicx}

\newcommand{\Starname}{\mbox{LB-1}}

\def\kms{km\,s$^{-1}$}

\def\hea{He\,{\sc i}}

\usepackage{txfonts}

\begin{document}

   \title{The 'hidden' companion in LB-1 unveiled by spectral disentangling\thanks{Based on observations made with the Mercator Telescope, operated on the island of La Palma by the Flemish Community, at the Spanish Observatorio del Roque de los Muchachos of the Instituto de Astrofísica de Canarias, and on observations made with the FEROS spectrograph attached to the 2.2-m MPG/ESO telescope at the La Silla observatory}$^{,}$\thanks{HERMES data are are only available at the CDS via anonymous ftp
to cdsarc.u-strasbg.fr (130.79.128.5) or via \url{http://cdsarc.u-strasbg.fr/viz-bin/cat/J/A+A/639/L6}} 
   }
   \author{
   T.\ Shenar, J.\ Bodensteiner,
   M.\ Abdul-Masih, M.\ Fabry, L.\ Mahy, P.\ Marchant, 
   \\ G.\ Banyard,  D.\ M.\ Bowman, K.\ Dsilva, C.\ Hawcroft, M.\ Reggiani, H.\ Sana
   }

   \institute{Institute of Astronomy, KU Leuven, Celestijnenlaan 200D, B-3001 Leuven, Belgium \\
              \email{tomer.shenar@kuleuven.be}
    }

   \date{Received Month xx, xxxx; accepted Month xx, xxxx}


  \abstract
   { The intriguing binary \object{LS V +22 25} (\Starname) has drawn much attention following claims of it being a single-lined spectroscopic binary with a 79-day orbit comprising a B-type star and a ${\approx}\,70\,M_\odot$ black hole -- the most massive stellar black hole reported to date. Subsequent studies demonstrated a lack of evidence for a companion of such great mass. Recent analyses have implied that the primary star is a stripped He-rich star with peculiar sub-solar abundances of heavy elements, such as Mg and Fe. However, the nature of the secondary, which was proposed to be a black hole, a neutron star, or a main sequence star, remains unknown.
   }
   {
   Based on 26 newly acquired spectroscopic observations secured with the HERMES and FEROS spectrographs  covering the orbit of the system, we perform an orbital analysis and spectral disentangling of \Starname~to elucidate the nature of the system. 
   }
   {To derive the radial velocity semi-amplitude $K_2$ of the secondary and extract the spectra of the two components,
   we used two independent disentangling methods: the shift-and-add technique and Fourier disentangling with FDBinary. We used atmosphere models to constrain the surface properties and abundances.
   }
   { 
   Our disentangling and spectral analysis shows that \Starname~contains two components of comparable brightness in the optical.  The narrow-lined primary, which we estimate to contribute $\approx55$\% in the optical, has spectral properties that suggest that it is a stripped star:  it has a small spectroscopic mass ($\approx 1\,M_\odot$) for a B-type star and it is He- and N-rich. Unlike previous reports, the abundances of heavy elements are found to be solar.   The 'hidden' secondary, which contributes about 45\% of the optical flux, is a rapidly rotating ($\varv \sin i\approx300\,$\kms) B3~V star with a decretion disk -- a Be star.  As a result of its rapid rotation and dilution, the photospheric absorption lines of the secondary are not readily apparent in the individual observations.  We measure a semi-amplitude for this star of $K_2=11.2{\pm}1.0\,$\kms~and adopting a mass of
   $M_2=7\pm2\,M_\odot$ typical for B3~V stars, we derive an orbital mass for the stripped primary of $M_1=1.5\pm0.4\,M_\odot$. The orbital inclination of $39{\pm}4^\circ$ implies a near-critical rotation for the Be secondary  ($\varv_{\rm eq}\approx470\,$\kms).
   }
   {
   \Starname~does not contain a compact object. Instead, it is a rare Be binary system consisting of  a stripped star (the former mass donor) and a  Be star rotating at near its critical velocity (the former mass accretor). This system is a clear example that binary interactions play a decisive role in the production of rapid stellar rotators and Be stars.
   } 

   \authorrunning{Shenar et al.}

   \maketitle
%

%
\section{Introduction}\label{Sec:intro}

The extraordinary binary system \object{LS V +22 25} (\object{\Starname} hereafter) was reported by  \citet{Liu2019} to comprise a B-type star (the primary) in a \mbox{79-day} orbit with a ${\approx}\,70\,M_\odot$ black hole (BH). This BH mass significantly exceeded mass measurements of massive stellar BHs, including the record-breaking measurements from gravitational-wave events \citep{Abbott2019}.  The formation of such a massive BH, especially in solar metallicity environments, challenges our current understanding of stellar evolution, a problem that has been addressed by a series of recent studies \citep{Groh2019, Eldridge2019, Safarzadeh2019, Belczynski2020}. 

The mass measurement of the BH by \citet{Liu2019} relies on the derived orbit and spectral properties of the primary and on radial velocity (RV) measurements of the wings of the H$\alpha$ emission line.  As \citet{Liu2019} notes, the H$\alpha$ line appears to exhibit a low-amplitude, anti-phase motion  relative  to  the  primary  star, leading the authors to speculate that it originates in an accretion disk around a ${\approx}\,70\,M_\odot$ BH secondary.
However, \citet{AbdulMasihBH2020} and \citet{ElBadry2020} have since independently shown that this anti-phase signature can be explained as a spurious result that emerges from the motion of the broad absorption wings of the B-type primary on top of a static H$\alpha$ line.
These results cast doubt on the measurement of the large mass of the secondary from \citet{Liu2019}. The derived binary mass function of $1.2\,M_\odot$ sets a definite lower limit on the mass of the  unseen secondary. However, its nature remains unknown.

\citet{Simon-Diaz2020} performed a spectroscopic analysis of \Starname~while relaxing the assumption of local thermodynamic equilibrium (non-LTE). They classified the primary star as B6~IV and derived significantly lower values for the effective temperature ($T_{\rm eff}{=}14\,$kK) and surface gravity ($\log g{=}3.5\,$[cgs]) compared with  \citet{Liu2019}. Similar values were reported from an LTE analysis performed by \citet{AbdulMasihBH2020}.

Even lower values of $T_{\rm eff}{=}12.7\,$kK and $\log g{=}3.0\,$[cgs] were recently reported by \citet{Irrgang2020}, who utilised hybrid LTE and non-LTE models. They suggested that the unseen secondary component is a neutron star  or a faint main-sequence star and that the primary star is a stripped helium star that lost its \mbox{H-rich} envelope due to a past binary mass-transfer event. This claim was based on anomalous abundances derived for the star, consistent with processed material from the CNO cycle (enriched N, depleted C and O), also confirmed by \citet{Simon-Diaz2020}. 
A notable peculiarity that was consistently reported in the aforementioned studies is the sub-solar abundance derived for several heavy elements such as Mg, Si, and Fe, which are not expected to change throughout the evolution of the star. This led \citet{AbdulMasihBH2020} to speculate that the secondary may be a rapidly rotating star that dilutes some of the primary spectral lines, leading to an apparent sub-solar metal content.

In this letter, we present firm evidence that the  unseen secondary is not a compact object, but a rapidly-rotating Be star, and that the properties of the primary star in \Starname~are indeed consistent with it being a stripped star. We show this through the spectral disentangling and analysis  of 26 recently obtained optical spectra 
that cover the full 79\,d orbit.


\section{Observations and data reduction}
\label{sec:ObsDat}
From December 2019 to March 2020, we obtained 24 epochs of observations with the {\sc HERMES} spectrograph mounted on the Mercator telescope \citep{Raskin2011} that uniformly sample the orbit of \Starname. HERMES spectra cover the wavelength range from 3770 to 9000\,$\AA$ with a spectral resolving power of $\sim$\,85\,000. Additionally, we obtained two epochs with the FEROS spectrograph \citep{Kaufer1999} mounted on the  MPG/ESO 2.2-m telescope at the La Silla observatory. The wavelength coverage of FEROS ranges from 3500 to 9200 $\AA$ with a spectral resolving power of 48\,000. The journal of the observations, including typical signal-to-noise ratios (S/N), is given in Table\,\ref{tab:obs_log}.

Standard calibrations, including bias and flat-field corrections and wavelength calibrations, were performed for both data sets using their respective pipelines. Barycentric correction was applied. On some of the nights, multiple observations were taken consecutively (see Table\,\ref{tab:obs_log}). Concomitant spectra obtained on the same night were combined using a S/N-weighted average. The spectra were individually normalized by fitting a spline through anchor points in the continuum of each spectrum.

\section{Orbital analysis} \label{sec:orbitanalysis}

\subsection{The orbit of the primary}\label{subsec:primorbit}
We measured the RVs of the primary in the 26 available spectra using the method described in \citet{Sana2013}, which relies on Gaussian fitting to the spectral lines simultaneously at all observing epochs.  
Here, we used three high-S/N He\,{\sc i} lines at $\lambda \lambda$ 4713, 5015, and 6678 $\AA$.

We subsequently fitted an orbit to the RV data (Fig.\,\ref{fig:orbit}) using the {\sc Python} package \textsc{spinOS}\footnote{https://github.com/roeiboot4/spinOS}. Given that the derived eccentricity of the orbit ($e{=}0.0036\pm 0.0021$) is not significant \citep{LucySweeney71}, we adopted a circular orbit. As our orbital solution (root-mean-square; rms = 0.86\,\kms) is in good agreement with the orbital solution presented in \citet{Liu2019}, we joined the two data sets, which, when combined, gave us a time base of 4\,yr, yielding an unweighted rms of 2.7\,\kms. 
The best-fit parameters using the combined data set are included in Table\,\ref{tab:Bstar_properties}.

\begin{table}[]\centering
    \caption{ {\it Upper:} Orbital parameters derived based on our orbital analysis and grid disentanglement, along with their 1$\sigma$ errors.  {\it Lower:} Estimated physical parameters of the components of \Starname~based on a spectral analysis with GSSP and PoWR.}
    \begin{tabular}{llcc} \hline \hline
        Parameter & primary & secondary \\ 
        \hline
        Spectral type    & stripped star & B3~Ve  \\     
         $\mathrm{P}_\mathrm{orb}$\,[d] & \multicolumn{2}{c}{$78.7999 \pm 0.0097$} \\
         T$_0$\,[HJD] ($\phi=0$) & \multicolumn{2}{c}{$2458845.54 \pm 0.03$} \\ 
         e  & \multicolumn{2}{c}{0 (fixed)} \\
         $\gamma$\,[\kms] &  \multicolumn{2}{c}{28.74 $\pm$ 0.09} \\
         $K$\,[\kms] &  52.94 $\pm$ 0.13 & $11.2 \pm 1.0$ \\
         $M\,\sin^3 i$\,[$M_\odot$] &  $0.38 \pm 0.04$ & $1.78 \pm 0.03$ \\
         $a\,\sin i$\,[$R_\odot$] &  $82.6 \pm 0.2$ & $17.5\ \pm 1.6$ \\      
         $q(\frac{M_2}{M_1})$  &   \multicolumn{2}{c}{$4.7 \pm 0.4$} \\
         $M_{\rm dyn}$\,[$M_\odot$] &  $1.5 \pm 0.4$ & $7 \pm 2$ (fixed) \\   
         $i$\,[deg] &  \multicolumn{2}{c}{39$ \pm $4}\\
         \hline
        $T_{\rm eff}$\,[kK] & 12.7 & 18   \\
        $\log g$\,[cgs] & 3.0 & 4.0   \\
        flux $f/f_{\rm tot}(V)$ & 0.55 (fixed)  & 0.45 (fixed)  \\
        $\log\,L\,[L_\odot]$  & 2.8  & 3.1  \\ 
        $R\,[R_\odot]$ & 5.4  & 3.7  \\         
        $M_{\rm spec}$\,[$M_\odot$] & 1.1  & 5  \\              
        $\varv\,\sin i$\,[\kms] & 7 $\pm$ 2 & $300 \pm 50$ \\
        $\varv_\mathrm{macro}$\,[\kms] & 4 $\pm$ 2 & 30 (fixed) \\
        $\varv_\mathrm{micro}$\,[\kms] & 2 (fixed) & 15 (fixed) \\
        $Z/Z_\odot$ &  1  & 1 \\ 
        n(He)/n(H)  & 0.21 & solar (0.08) \\
    \hline
    \end{tabular}
    \label{tab:Bstar_properties}
\end{table}

\subsection{Evidence for Balmer emission tracing the primary}\label{subsec:trailing}

The morphologies of the Balmer emission lines in \Starname~are complex and variable.
\citet{Liu2019} suggested that the H$\alpha$ line comprises two contributions: a broad H$\alpha$ emission that moves with the putative BH secondary in anti-phase to H$\alpha$ absorption from the primary.  Here, we show that the primary  appears to exhibit H$\alpha$ emission, and not absorption. 

To illustrate this, Fig.\,\ref{fig:GrayScale} shows dynamic spectra of the H$\alpha$ and He\,{\sc i}\,$\lambda 6678$ lines, demonstrating a complex emission pattern that traces the orbit of the primary. Figure\,\ref{fig:GrayScale} also shows two HERMES spectra taken close to quadrature, in which the narrow H$\alpha$ emission is shown to follow the orbit of the primary.
This is in contrast to the behaviour of the H$\alpha$-line wings, which appear to exhibit a slight anti-phase motion. However, it is not readily clear if this anti-phase motion contains a reflex motion of the secondary \citep{AbdulMasihBH2020, ElBadry2020}.


\section{Spectral disentangling: unveiling the companion}\label{sec:disen}

\begin{figure*}
    \centering
    \includegraphics[width=\textwidth]{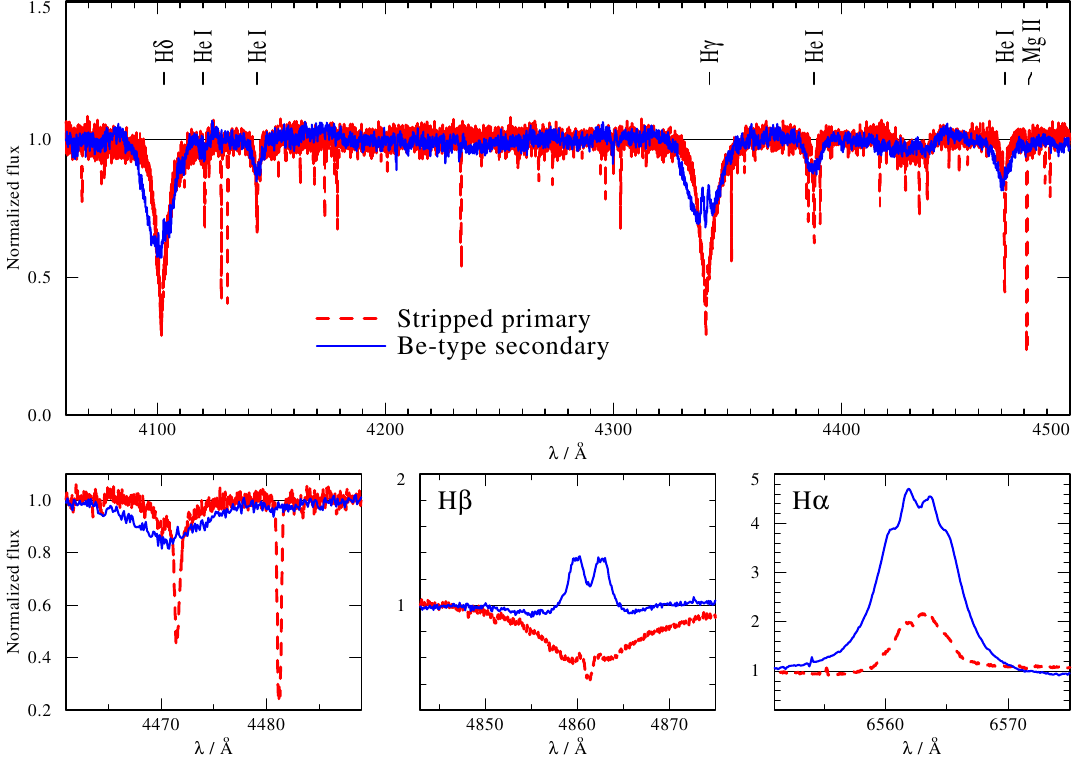}
    \caption{Disentangled spectra obtained with the shift-and-add technique. The spectrum of the Be secondary is binned at $\Delta \lambda =0.1\,$\AA~for clarity. The disentangled spectra obtained with FDBinary (Fourier disentangling) are identical within the S/N. The spectra are scaled assuming the light ratios 55\% and 45\% for the primary and secondary, respectively, which is justified in  Sect.\,\ref{subsec:Bprim}.}
    \label{fig:ShiftnAdd}
\end{figure*}


The fact that in previous studies there was no companion to the primary star to be readily seen in the spectra was interpreted as evidence for the presence of a dark companion (i.e. neutron star or BH). However, within the S/N levels, it is possible that a combination of rapid rotation and low light contribution could push a non-degenerate stellar companion below the detection threshold \citep{AbdulMasihBH2020}.


To extract the spectra of the individual components in \Starname~from the observations, we utilise spectral disentangling 
using two different methods: the shift-and-add technique in wavelength space and disentangling in Fourier space. These methods are described below and the comparison of their results offers a consistency check of the robustness of the disentangling process. 
As used here, both methods rely on the orbital parameters of the system. Since the RV amplitude of the secondary object is not constrained, we fix all orbital parameters and perform a disentangling along the $K_2$-axis. We infer $K_2$ by minimising the  reduced $\chi^2(K_2)$. The scaling of the final disentangled spectra depends on the light contribution of each component (see Appendix \,\ref{sec:SpTanalysis}). However, the $\chi^2$ statistic is independent of the light ratio.



\subsection{Disentangling using shift-and-add}\label{subsubsec:shiftnadd}

Shift-and-add is a widely used disentangling technique \citep[e.g.][]{Marchenko1998, Gonzalez2006, Mahy2012, Shenar2018}. It is an iterative procedure that uses the disentangled spectra obtained in the $j^\mathrm{th}$ iteration, $A_j$ and $B_j$, to calculate the disentangled spectra for the $j+1^\mathrm{th}$ iteration.


Fixing the orbital parameters (Table\,\ref{tab:Bstar_properties}), 
we evaluate the reduced $\chi^2(K_2)$ for a series of $K_2$ values in the range of 0 to 100\,\kms~ in steps of $\Delta K_2 = 0.5\,$\kms.  We consider different wavelength ranges:  the full spectrum, He\,{\sc i} lines, and Balmer lines. The H$\alpha$ line implies a relatively well-defined minimum at $K_2\,{\approx}\,10\,$\kms. The H$\beta$, H$\gamma$, and H$\delta$ lines imply similar values, but due to their much lower S/N, their minima are poorly localised (Fig.\,\ref{fig:Habg}).
A parabola fit to the minimum region in the combined reduced $\chi^2$ of all Balmer lines yields  \mbox{$K_2{=} 11.3\pm1.4$\,\kms}, where the $1\sigma$ error is calculated from the corresponding $\chi^2(K_2)$ contour. The disentangled spectra are shown in Fig.\,\ref{fig:ShiftnAdd}.



\subsection{Fourier disentangling with FDBinary}\label{subsubsec:Fourier}


Unlike the shift-and-add technique, which works in wavelength space, the FDBinary disentangling tool operates in Fourier space. The method was developed by \citet{Hadrava1995} and implemented in FDBinary by  \citet{Ilijic2004}, where a detailed description can be found.

Using the orbital parameters listed in Table \ref{tab:Bstar_properties}, we perform a $\chi^2$ minimisation by varying $K_2$. Errors are estimated by running 3000 Monte Carlo samples of the minimisation process, while altering the spectra and the orbit by adding white noise. Our grid ranges from 0.5 to 25\,\kms\ in steps of 0.5\,\kms\ for $K_2$. Instead of disentangling the entire wavelength range, we focus specifically on Balmer and \hea\, lines. We find that the minimal $\chi^2$ distance is achieved for $K_2 = 11.2 \pm 1.0$\,\kms~(1$\sigma$ confidence interval).

\subsection{Disentangled spectra of the two components}\label{subsec:disspec}

The results obtained by the two different techniques in Sects.\,\ref{subsubsec:shiftnadd} and \ref{subsubsec:Fourier}  agree well. 
The derived semi-amplitudes $K_2$  are consistent within the $1\sigma$ errors and the spectra for the two components derived from both methods are consistent within the S/N. The derived $K_2$ value of 11\,\kms~also  agrees with the recent independent measurement provided by \citet{Liu2020}.


As is evident from Fig.\,\ref{fig:ShiftnAdd}, our algorithm successfully separates the observed spectra into two stellar components. The first exhibits a multitude of narrow lines and is clearly associated with the  primary star seen in the individual observations. We note the H$\alpha$ and H$\beta$ emission features observed for the primary (see Sect.\,\ref{subsec:trailing}). In contrast, the secondary component exhibits significantly broadened lines belonging primarily to the Balmer series and to He\,{\sc i}. The Balmer lines portray double-peak emission profiles that are characteristic of classical Be stars: B-type stars possessing a decretion disk  \citep[e.g.][]{Rivinius2013}.

We provide a spectral classification and analysis of both components in Sect.\,\ref{sec:SpTanalysis}. The derived parameters are given in Table\,\ref{tab:Bstar_properties}.  The properties of the system  suggest that \Starname~is a Be binary that formed through a past mass-transfer event \citep{Pols1991, deMink2014}, currently comprising a stripped star (the mass donor) and a critically rotating Be star (the mass accretor).
In contrast to \citet{Liu2019}, we find no evidence for the presence of a compact object  in \Starname, nor do we find evidence that the Be component is a static tertiary, as has recently been proposed by \citet{Rivinius2020}.

\section{Conclusions} \label{sec:conclusion}

We performed an orbital analysis and spectral disentangling of the intriguing \Starname~binary system using newly-acquired spectroscopic observations that cover its 79-day orbit. We show that the binary does not contain a compact object, but rather, it consists of two non-degenerate components: a stripped primary and a rapidly rotating B3~Ve secondary. 

The orbital inclination and the mass of the primary can be  estimated by calibrating the mass of the B3~Ve secondary to a typical value for its spectral type and estimated parameters. Adopting $M_2{=}7{\pm}2\,M_\odot$ \citep{Cox2000} for the B3~Ve secondary, we obtain a narrow constraint on the inclination of $39{\pm}4^\circ$.   This, in turn, implies that the equatorial rotational velocity of the Be star is $\varv_\mathrm{eq}{\approx}470$\,\kms, close to critical \citep[cf.][]{Townsend2004}. The orbital mass of the stripped star is then $M_1 {=}1.5\pm0.4\,M_\odot$,  making it a potential core-collapse supernova progenitor \citep[e.g.][]{Zapartas2019}.

\Starname~thus represents a Be binary system in which the Be secondary has formed through a previous mass transfer event, having gained mass from the originally more massive primary star. This is a clear example of how binary interactions   act as an important agent in producing rapid stellar rotators and Be stars \citep[e.g.][]{Pols1991, deMink2013}. 

The low temperature of the stripped primary implies that it is thermally unstable, most likely contracting towards the He main sequence.  \Starname~is therefore a rare progenitor to Be binaries that host subdwarf stars, such as \object{$\phi$\,Per} \citep{Gies1998, Schootemeijer2018} and \object{o Pup} \citep{Koubsky2012}, potentially the most massive of its kind detected to date. These results pave the way to tailored binary evolution modeling of the \Starname~system, which will shed new light onto the evolutionary status of this extraordinary object and contribute to our understanding of binary-interaction processes.

\begin{acknowledgements}
Based on observations obtained with the HERMES spectrograph, which is supported by the Research Foundation - Flanders (FWO), Belgium, the Research Council of KU Leuven, Belgium, the Fonds National de la Recherche Scientifique (F.R.S.-FNRS), Belgium, the Royal Observatory of Belgium, the Observatoire de Genève, Switzerland and the Thüringer Landessternwarte Tautenburg, Germany. The authors gratefully acknowledge MPIA, Heidelberg for access to the FEROS spectrograph for use in this work.
The research leading to these results has received funding from the European
Research Council (ERC) under the European Union’s Horizon 2020 research
and innovation programme (grant agreement numbers 772225: MULTIPLES and
670519: MAMSIE), and the FWO Odysseus program under project G0F8H6N. TS acknowledges helpful discussions with Helge Todt and Avishai Gilkis.
\end{acknowledgements}

\bibliographystyle{aa}
\bibliography{papers}

\begin{thebibliography}{42}
\expandafter\ifx\csname natexlab\endcsname\relax\def\natexlab#1{#1}\fi

\bibitem[{{Abbott} {et~al.}(2019){Abbott}, {Abbott}, {Abbott}, {Abraham},
  {Acernese}, {Ackley}, {Adams}, {Adhikari}, {Adya}, {Affeldt}, {Agathos},
  {Agatsuma}, {Aggarwal}, {Aguiar}, {Aiello}, {Ain}, {Ajith}, {Allen},
  {Allocca}, {Aloy}, {Altin}, {Amato}, {Ananyeva}, {Anderson}, {Anderson},
  {Angelova}, {Antier}, {Appert}, {Arai}, {Araya}, {Areeda}, {Ar{\`e}ne},
  {Arnaud}, {Arun}, {Ascenzi}, {Ashton}, {Aston}, {Astone}, {Aubin}, {Aufmuth},
  {AultONeal}, {Austin}, {Avendano}, {Avila-Alvarez}, {Babak}, {Bacon},
  {Badaracco}, {Bader}, {Bae}, {Baker}, {Baldaccini}, {Ballardin}, {Ballmer},
  {Banagiri}, {Barayoga}, {Barclay}, {Barish}, {Barker}, {Barkett}, {Barnum},
  {Barone}, {Barr}, {Barsotti}, {Barsuglia}, {Barta}, {Bartlett}, {Bartos},
  {Bassiri}, {Basti}, {Bawaj}, {Bayley}, {Bazzan}, {B{\'e}csy}, {Bejger},
  {Belahcene}, {Bell}, {Beniwal}, {Berger}, {Bergmann}, {Bernuzzi}, {Bero},
  {Berry}, {Bersanetti}, {Bertolini}, {Betzwieser}, {Bhand are}, {Bidler},
  {Bilenko}, {Bilgili}, {Billingsley}, {Birch}, {Birney}, {Birnholtz},
  {Biscans}, {Biscoveanu}, {Bisht}, {Bitossi}, {Bizouard}, {Blackburn},
  {Blackman}, {Blair}, {Blair}, {Blair}, {Bloemen}, {Bode}, {Boer}, {Boetzel},
  {Bogaert}, {Zucker}, {Zweizig}, {LIGO Scientific Collaboration}, \& {Virgo
  Collaboration}}]{Abbott2019}
{Abbott}, B.~P., {Abbott}, R., {Abbott}, T.~D., {et~al.} 2019, Physical Review
  X, 9, 031040

\bibitem[{{Abdul-Masih} {et~al.}(2020){Abdul-Masih}, {Banyard}, {Bodensteiner},
  {Bordier}, {Bowman}, {Dsilva}, {Fabry}, {Hawcroft}, {Mahy}, {Marchant},
  {Raskin}, {Reggiani}, {Shenar}, {Tkachenko}, {Van Winckel}, {Vermeylen}, \&
  {Sana}}]{AbdulMasihBH2020}
{Abdul-Masih}, M., {Banyard}, G., {Bodensteiner}, J., {et~al.} 2020, \nat, 580,
  E11

\bibitem[{{Bailer-Jones} {et~al.}(2018){Bailer-Jones}, {Rybizki}, {Fouesneau},
  {Mantelet}, \& {Andrae}}]{Bailer-Jones2018}
{Bailer-Jones}, C.~A.~L., {Rybizki}, J., {Fouesneau}, M., {Mantelet}, G., \&
  {Andrae}, R. 2018, \aj, 156, 58

\bibitem[{{Belczynski} {et~al.}(2020){Belczynski}, {Hirschi}, {Kaiser}, {Liu},
  {Casares}, {Lu}, {O'Shaughnessy}, {Heger}, {Justham}, \&
  {Soria}}]{Belczynski2020}
{Belczynski}, K., {Hirschi}, R., {Kaiser}, E.~A., {et~al.} 2020, \apj, 890, 113

\bibitem[{{Cox}(2000)}]{Cox2000}
{Cox}, A.~N. 2000, {Allen's astrophysical quantities}

\bibitem[{{de Mink} {et~al.}(2013){de Mink}, {Langer}, {Izzard}, {Sana}, \& {de
  Koter}}]{deMink2013}
{de Mink}, S.~E., {Langer}, N., {Izzard}, R.~G., {Sana}, H., \& {de Koter}, A.
  2013, \apj, 764, 166

\bibitem[{{de Mink} {et~al.}(2014){de Mink}, {Sana}, {Langer}, {Izzard}, \&
  {Schneider}}]{deMink2014}
{de Mink}, S.~E., {Sana}, H., {Langer}, N., {Izzard}, R.~G., \& {Schneider},
  F.~R.~N. 2014, \apj, 782, 7

\bibitem[{{El-Badry} \& {Quataert}(2020)}]{ElBadry2020}
{El-Badry}, K. \& {Quataert}, E. 2020, \mnras, 493, L22

\bibitem[{{Eldridge} {et~al.}(2019){Eldridge}, {Stanway}, {Breivik}, {Casey},
  {Steeghs}, \& {Stevance}}]{Eldridge2019}
{Eldridge}, J.~J., {Stanway}, E.~R., {Breivik}, K., {et~al.} 2019, arXiv
  e-prints, arXiv:1912.03599

\bibitem[{{Gies} {et~al.}(1998){Gies}, {Bagnuolo}, {Ferrara}, {Kaye},
  {Thaller}, {Penny}, \& {Peters}}]{Gies1998}
{Gies}, D.~R., {Bagnuolo}, William~G., J., {Ferrara}, E.~C., {et~al.} 1998,
  \apj, 493, 440

\bibitem[{{Gonz{\'a}lez} \& {Levato}(2006)}]{Gonzalez2006}
{Gonz{\'a}lez}, J.~F. \& {Levato}, H. 2006, \aap, 448, 283

\bibitem[{{Groh} {et~al.}(2019){Groh}, {Farrell}, {Meynet}, {Smith}, {Murphy},
  \& {Allan}}]{Groh2019}
{Groh}, J.~H., {Farrell}, E., {Meynet}, G., {et~al.} 2019, arXiv e-prints,
  arXiv:1912.00994

\bibitem[{{Guetter}(1968)}]{Guetter1968}
{Guetter}, H.~H. 1968, \pasp, 80, 197

\bibitem[{{Hadrava}(1995)}]{Hadrava1995}
{Hadrava}, P. 1995, \aaps, 114, 393

\bibitem[{{Hainich} {et~al.}(2019){Hainich}, {Ramachandran}, {Shenar}, {Sand
  er}, {Todt}, {Gruner}, {Oskinova}, \& {Hamann}}]{Hainich2019}
{Hainich}, R., {Ramachandran}, V., {Shenar}, T., {et~al.} 2019, \aap, 621, A85

\bibitem[{{Hamann} \& {Gr{\"a}fener}(2003)}]{Hamann2003}
{Hamann}, W.~R. \& {Gr{\"a}fener}, G. 2003, \aap, 410, 993

\bibitem[{{Herman} {et~al.}(1959){Herman}, {Barin}, \& {Pendzel}}]{Herman1959}
{Herman}, R., {Barin}, M.~T., \& {Pendzel}, M. 1959, Annales d'Astrophysique,
  22, 540

\bibitem[{{Ilijic} {et~al.}(2004){Ilijic}, {Hensberge}, {Pavlovski}, \&
  {Freyhammer}}]{Ilijic2004}
{Ilijic}, S., {Hensberge}, H., {Pavlovski}, K., \& {Freyhammer}, L.~M. 2004,
  Astronomical Society of the Pacific Conference Series, Vol. 318, {Obtaining
  normalised component spectra with FDBinary}, ed. R.~W. {Hilditch},
  H.~{Hensberge}, \& K.~{Pavlovski}, 111--113

\bibitem[{{Irrgang} {et~al.}(2020){Irrgang}, {Geier}, {Kreuzer}, {Pelisoli}, \&
  {Heber}}]{Irrgang2020}
{Irrgang}, A., {Geier}, S., {Kreuzer}, S., {Pelisoli}, I., \& {Heber}, U. 2020,
  \aap, 633, L5

\bibitem[{{Kaufer} {et~al.}(1999){Kaufer}, {Stahl}, {Tubbesing},
  {N{\o}rregaard}, {Avila}, {Francois}, {Pasquini}, \& {Pizzella}}]{Kaufer1999}
{Kaufer}, A., {Stahl}, O., {Tubbesing}, S., {et~al.} 1999, The Messenger, 95, 8

\bibitem[{{Koubsk{\'y}} {et~al.}(2012){Koubsk{\'y}}, {Kotkov{\'a}}, {Votruba},
  {{\v{S}}lechta}, \& {Dvo{\v{r}}{\'a}kov{\'a}}}]{Koubsky2012}
{Koubsk{\'y}}, P., {Kotkov{\'a}}, L., {Votruba}, V., {{\v{S}}lechta}, M., \&
  {Dvo{\v{r}}{\'a}kov{\'a}}, {\v{S}}. 2012, \aap, 545, A121

\bibitem[{{Liu} {et~al.}(2019){Liu}, {Zhang}, {Howard}, {Bai}, {Lu}, {Soria},
  {Justham}, {Li}, {Zheng}, {Wang}, {Belczynski}, {Casares}, {Zhang}, {Yuan},
  {Dong}, {Lei}, {Isaacson}, {Wang}, {Bai}, {Shao}, {Gao}, {Wang}, {Niu},
  {Cui}, {Zheng}, {Mu}, {Zhang}, {Wang}, {Heger}, {Qi}, {Liao}, {Lattanzi},
  {Gu}, {Wang}, {Wu}, {Shao}, {Shen}, {Wang}, {Bregman}, {Di Stefano}, {Liu},
  {Han}, {Zhang}, {Wang}, {Ren}, {Zhang}, {Zhang}, {Wang}, {Cabrera-Lavers},
  {Corradi}, {Rebolo}, {Zhao}, {Zhao}, {Chu}, \& {Cui}}]{Liu2019}
{Liu}, J., {Zhang}, H., {Howard}, A.~W., {et~al.} 2019, \nat, 575, 618

\bibitem[{{Liu} {et~al.}(2020){Liu}, {Zheng}, {Soria}, {Aceituno}, {Zhang},
  {Lu}, {Wang}, {Hamann}, {Oskinova}, {Ramachandran}, {Yuan}, {Bai}, {Wang},
  {McKee}, {Wu}, {Wang}, {Lattanzi}, {Belczynski}, {Casares}, {Simon-Diaz},
  {Gonz{\'a}lez Hern{\'a}ndez}, \& {Rebolo}}]{Liu2020}
{Liu}, J., {Zheng}, Z., {Soria}, R., {et~al.} 2020, arXiv e-prints,
  arXiv:2005.12595

\bibitem[{{Lucy} \& {Sweeney}(1971)}]{LucySweeney71}
{Lucy}, L.~B. \& {Sweeney}, M.~A. 1971, \aj, 76, 544

\bibitem[{{Mahy} {et~al.}(2012){Mahy}, {Gosset}, {Sana}, {Damerdji}, {De
  Becker}, {Rauw}, \& {Nitschelm}}]{Mahy2012}
{Mahy}, L., {Gosset}, E., {Sana}, H., {et~al.} 2012, \aap, 540, A97

\bibitem[{{Marchenko} {et~al.}(1998){Marchenko}, {Moffat}, \&
  {Eenens}}]{Marchenko1998}
{Marchenko}, S.~V., {Moffat}, A. F.~J., \& {Eenens}, P. R.~J. 1998, \pasp, 110,
  1416

\bibitem[{Pols {et~al.}(1991)Pols, Cote, Waters, \& Heise}]{Pols1991}
Pols, O., Cote, J., Waters, L. B. F.~M., \& Heise, J. 1991, A{\&}A, 241, 419

\bibitem[{Raskin {et~al.}(2011)Raskin, {Van Winckel}, Hensberge, Jorissen,
  Lehmann, Waelkens, Avila, {De Cuyper}, Degroote, Dubosson, Dumortier, Fremat,
  Laux, Michaud, Morren, Padilla, Pessemier, Prins, Smolders, {Van Eck}, \&
  Winkler}]{Raskin2011}
Raskin, G., {Van Winckel}, H., Hensberge, H., {et~al.} 2011, A{\&}A, 526, A69

\bibitem[{{Rivinius} {et~al.}(2020){Rivinius}, {Baade}, {Hadrava}, {Heida}, \&
  {Klement}}]{Rivinius2020}
{Rivinius}, T., {Baade}, D., {Hadrava}, P., {Heida}, M., \& {Klement}, R. 2020,
  \aap, 637, L3

\bibitem[{{Rivinius} {et~al.}(2013){Rivinius}, {Carciofi}, \&
  {Martayan}}]{Rivinius2013}
{Rivinius}, T., {Carciofi}, A.~C., \& {Martayan}, C. 2013, \aapr, 21, 69

\bibitem[{{Safarzadeh} {et~al.}(2019){Safarzadeh}, {Ramirez-Ruiz}, \&
  {Belczynski}}]{Safarzadeh2019}
{Safarzadeh}, M., {Ramirez-Ruiz}, E., \& {Belczynski}, K. 2019, arXiv e-prints,
  arXiv:1912.10456

\bibitem[{{Sana} {et~al.}(2013){Sana}, {de Koter}, {de Mink}, {Dunstall},
  {Evans}, {H{\'e}nault-Brunet}, {Ma{\'\i}z Apell{\'a}niz},
  {Ram{\'\i}rez-Agudelo}, {Taylor}, {Walborn}, {Clark}, {Crowther}, {Herrero},
  {Gieles}, {Langer}, {Lennon}, \& {Vink}}]{Sana2013}
{Sana}, H., {de Koter}, A., {de Mink}, S.~E., {et~al.} 2013, \aap, 550, A107

\bibitem[{{Sander} {et~al.}(2015){Sander}, {Shenar}, {Hainich},
  {G{\'\i}menez-Garc{\'\i}a}, {Todt}, \& {Hamann}}]{Sander2015}
{Sander}, A., {Shenar}, T., {Hainich}, R., {et~al.} 2015, \aap, 577, A13

\bibitem[{{Schootemeijer} {et~al.}(2018){Schootemeijer}, {G{\"o}tberg}, {de
  Mink}, {Gies}, \& {Zapartas}}]{Schootemeijer2018}
{Schootemeijer}, A., {G{\"o}tberg}, Y., {de Mink}, S.~E., {Gies}, D., \&
  {Zapartas}, E. 2018, \aap, 615, A30

\bibitem[{{Shenar} {et~al.}(2018){Shenar}, {Hainich}, {Todt}, {Moffat},
  {Sander}, {Oskinova}, {Ramachandran}, {Munoz}, {Pablo}, {Sana}, \&
  {Hamann}}]{Shenar2018}
{Shenar}, T., {Hainich}, R., {Todt}, H., {et~al.} 2018, \aap, 616, A103

\bibitem[{{Shulyak} {et~al.}(2004){Shulyak}, {Tsymbal}, {Ryabchikova},
  {St{\"u}tz}, \& {Weiss}}]{Shulyak2004}
{Shulyak}, D., {Tsymbal}, V., {Ryabchikova}, T., {St{\"u}tz}, C., \& {Weiss},
  W.~W. 2004, \aap, 428, 993

\bibitem[{{Sim{\'o}n-D{\'\i}az} {et~al.}(2020){Sim{\'o}n-D{\'\i}az}, {Ma{\'\i}z
  Apell{\'a}niz}, {Lennon}, {Gonz{\'a}lez Hern{\'a}ndez}, {Allende Prieto},
  {Castro}, {de Burgos}, {Dufton}, {Herrero}, {Toledo-Padr{\'o}n}, \&
  {Smartt}}]{Simon-Diaz2020}
{Sim{\'o}n-D{\'\i}az}, S., {Ma{\'\i}z Apell{\'a}niz}, J., {Lennon}, D.~J.,
  {et~al.} 2020, \aap, 634, L7

\bibitem[{{Slettebak}(1982)}]{Slettebak1982}
{Slettebak}, A. 1982, \apjs, 50, 55

\bibitem[{{Tkachenko}(2015)}]{Tkachenko2015}
{Tkachenko}, A. 2015, \aap, 581, A129

\bibitem[{Townsend {et~al.}(2004)Townsend, Owocki, \& Howarth}]{Townsend2004}
Townsend, R. H.~D., Owocki, S.~P., \& Howarth, I.~D. 2004, MNRAS, 350, 189

\bibitem[{{Tsymbal}(1996)}]{Tsymbal1996}
{Tsymbal}, V. 1996, Astronomical Society of the Pacific Conference Series, Vol.
  108, {STARSP: A Software System For the Analysis of the Spectra of Normal
  Stars}, ed. S.~J. {Adelman}, F.~{Kupka}, \& W.~W. {Weiss}, 198

\bibitem[{{Zapartas} {et~al.}(2019){Zapartas}, {de Mink}, {Justham}, {Smith},
  {de Koter}, {Renzo}, {Arcavi}, {Farmer}, {G{\"o}tberg}, \&
  {Toonen}}]{Zapartas2019}
{Zapartas}, E., {de Mink}, S.~E., {Justham}, S., {et~al.} 2019, \aap, 631, A5

\end{thebibliography}

\appendix

\section{Supplementary material}



\begin{figure}\centering
    \includegraphics[width=0.5\textwidth]{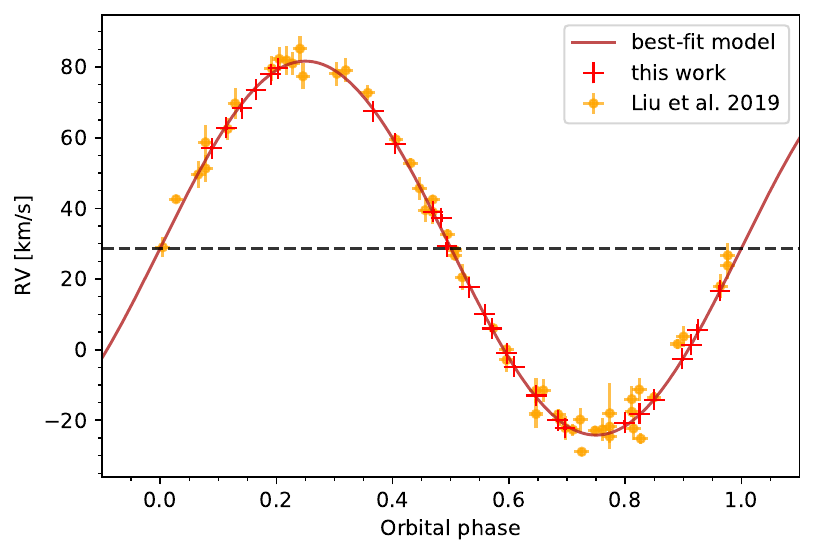}
    \caption{Best-fit orbital solution for a circular orbit overlaid with the RV measurements from HERMES and FEROS and the data from \citet{Liu2019}. For our data, the errors are smaller than the symbol size (see Table\,\ref{tab:obs_log}).}
    \label{fig:orbit}
\end{figure}

\begin{figure}
\centering
\begin{tabular}{c}
     \includegraphics[width=0.48\textwidth]{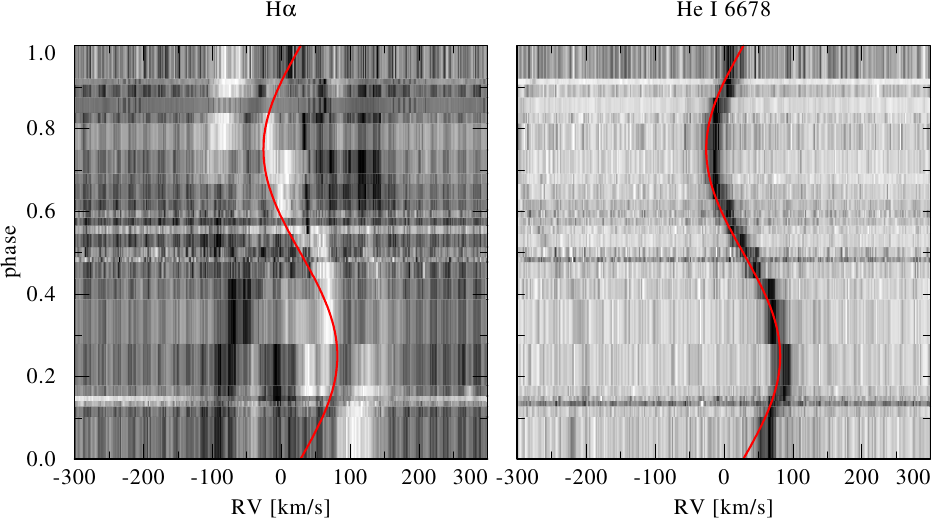} \\
     \includegraphics[width=0.49\textwidth]{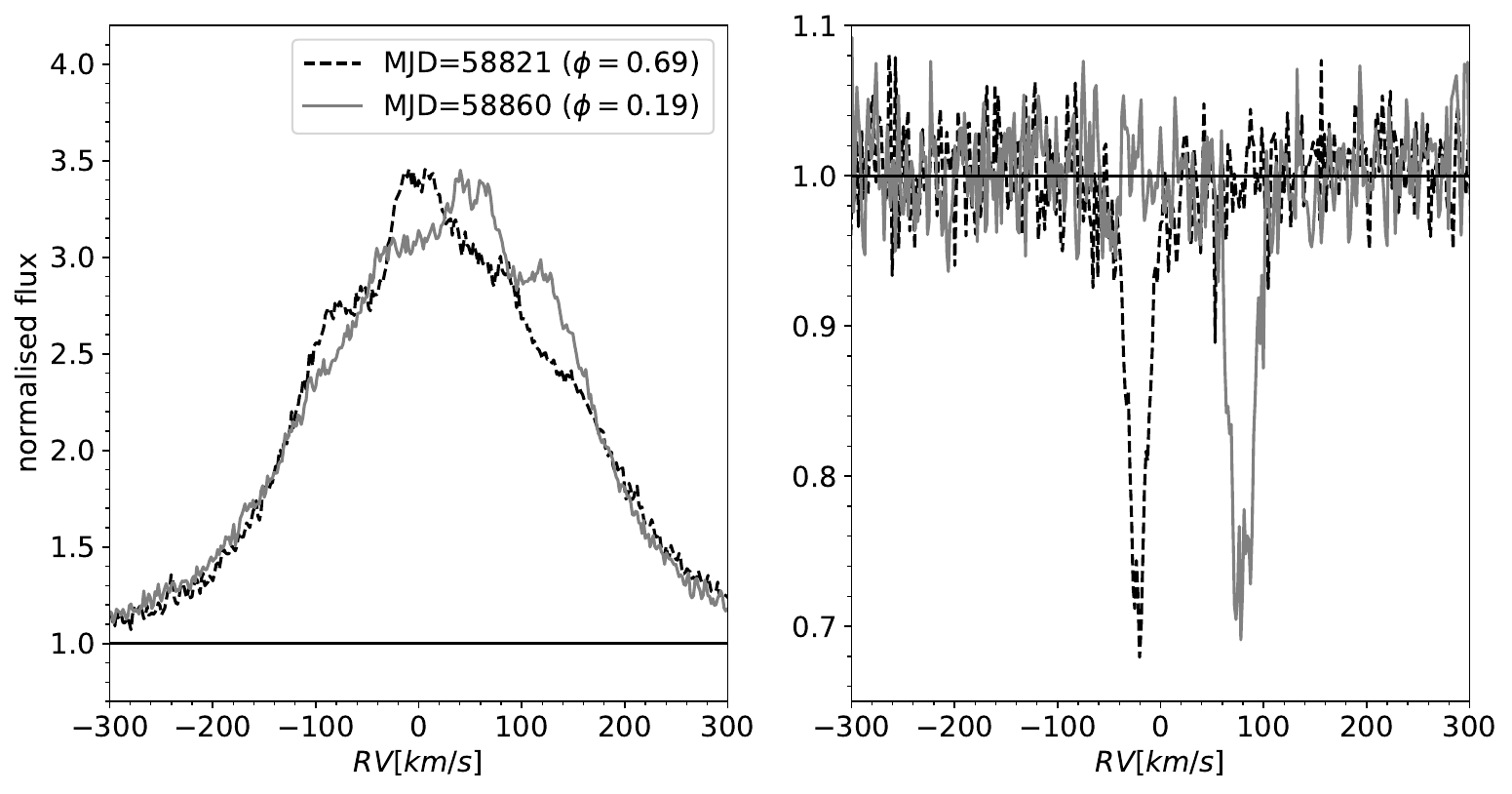} 
\end{tabular}
\caption{{\it Top:} Dynamical spectra of the H$\alpha$ (left) and He\,{\sc i}\,$\lambda 6678$ (right) lines in Doppler space phased using the ephemeris given in Table\,\ref{tab:Bstar_properties}, compared to the derived orbit (red line). The H$\alpha$ line is mean-subtracted to enhance the contrast. {\it Bottom:} Two HERMES spectra taken close to quadrature (see legend) of the H$\alpha$ and He\,{\sc i}\,$\lambda 6678$ lines, as in the top panel. The H$\alpha$ line exhibits a narrow emission peak on top that traces the orbit of the primary star.}
\label{fig:GrayScale}
\end{figure}


\begin{figure}\centering
    \includegraphics[width=0.5\textwidth]{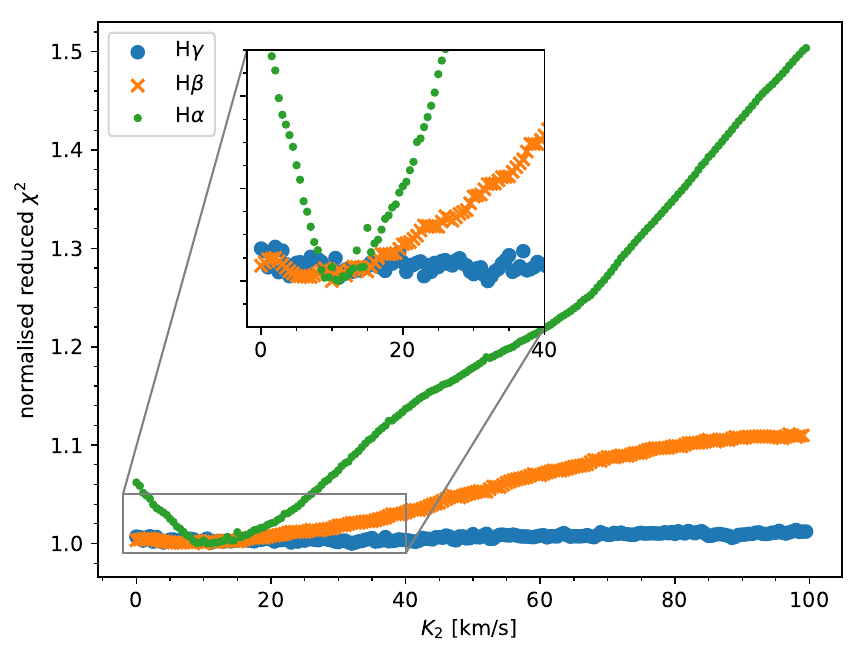}
    \caption{Reduced $\chi^2(K_2)$ obtained through shift-and-add grid disentangling for the H$\alpha$, H$\beta$, and H$\gamma$ lines (see legend), normalised to unity to allow for an easy comparison between the different curves.}
\label{fig:Habg}
\end{figure}

\begin{figure}\centering
    \includegraphics[width=0.5\textwidth]{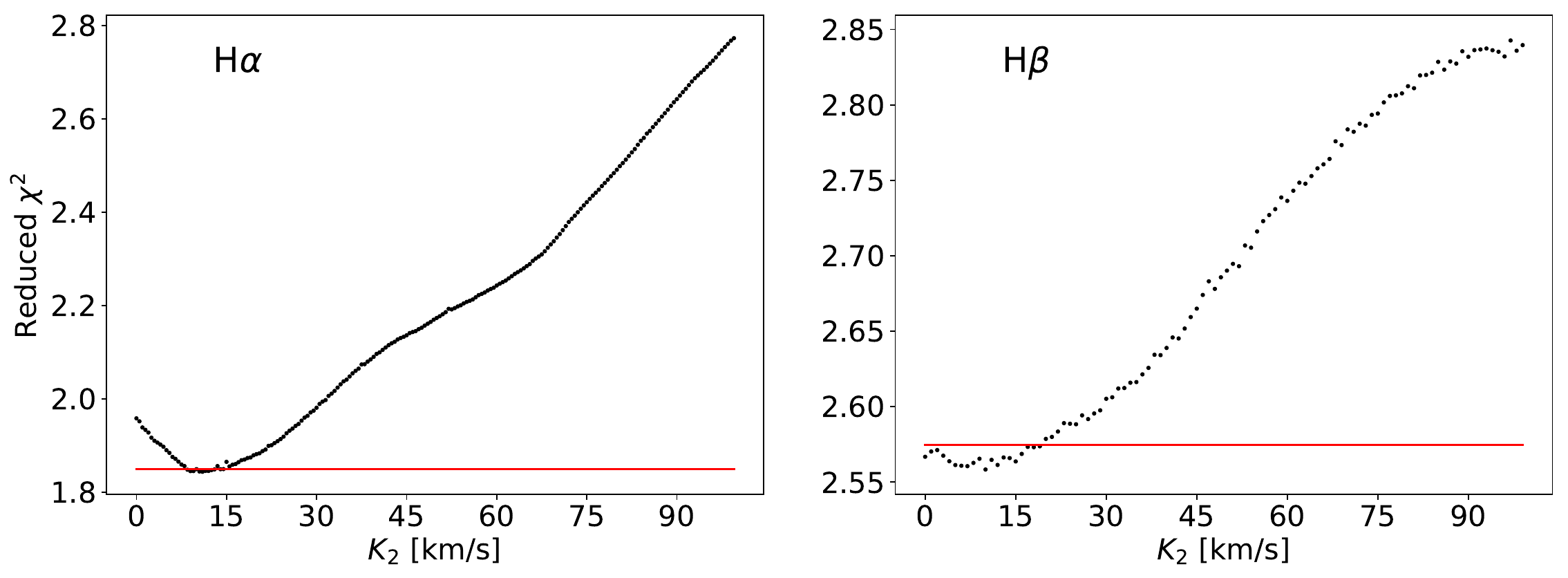}
    \caption{Absolute values of the reduced $\chi^2(K_2)$ for H$\alpha$ and H$\beta$. }
\label{fig:Habg_sep}
\end{figure}


\begin{table}[h!]\centering
    \caption{Journal of the observations of \Starname. The second column indicates whether the spectrum was taken with HERMES (H) or with FEROS (F). The third column shows the exposure time while the forth column gives the S/N ratio in the spectrum at $\approx 5400\,\AA$. The last column gives the RV of the primary star.}
    \label{tab:obs_log}
    \begin{tabular}{lllrr} \hline \hline 
     $\mathrm{HJD}$\,[d] & ins & time [s] & S/N & RV$_1$ [\kms]\\ \hline
2458821.593 & H & 2$\times$1800 & 37 & $-$22.19 $\pm$ 0.50 \\
2458833.719 & F & 1$\times$5400 & 93 & $-$14.12 $\pm$ 0.14 \\
2458837.576 & H & 2$\times$1800 & 33 &  $-$2.53 $\pm$ 0.54 \\
2458838.728 & F & 2$\times$2700 & 128 & 1.35 $\pm$ 0.10 \\
2458852.573 & H & 2$\times$1800 & 27 & 57.01 $\pm$ 0.60 \\
2458854.528 & H & 2$\times$1800 & 32 & 62.79 $\pm$ 0.60 \\
2458856.617 & H & 2$\times$1800 & 17 & 68.38 $\pm$ 1.20\\
2458858.534 & H & 3$\times$1800 & 27 & 73.58 $\pm$ 0.63\\
2458860.554 & H & 4$\times$1800 & 47 & 77.95 $\pm$ 0.44\\
2458861.544 & H & 2$\times$1800 & 32 & 79.62 $\pm$ 0.57 \\
2458874.496 & H & 4$\times$1800 & 38 & 67.46 $\pm$ 0.44\\
2458877.422 & H & 4$\times$1800 & 35 & 58.31 $\pm$ 0.52\\
2458882.525 & H & 2$\times$1800 & 31 & 39.07 $\pm$ 0.63\\
2458883.597 & H & 1$\times$2100 & 5 & 37.25 $\pm$ 1.20\\
2458884.460 & H & 1$\times$2400 & 17 & 29.30 $\pm$ 1.20\\
2458887.423 & H & 4$\times$1800 & 46 & 17.63 $\pm$ 0.43\\
2458889.527 & H & \vtop{\hbox{\strut 2$\times$1800 + }\hbox{\strut 2$\times$2000 }} & 38 & 10.04 $\pm$ 0.52\\
2458890.502 & H & 1$\times$1800 & 22 & 5.99 $\pm$ 0.86\\
2458892.492 & H & \vtop{\hbox{\strut 1$\times$1800 + }\hbox{\strut 1$\times$2000 }} & 22 & $-$0.97 $\pm$ 0.85 \\
2458893.496 & H & \vtop{\hbox{\strut 2$\times$1800 + }\hbox{\strut 1$\times$2000 }} & 38 & $-$4.78 $\pm$ 0.52\\
2458896.498 & H & 4$\times$1800 & 39 & $-$12.98 $\pm$ 0.52\\
2458899.434 & H & 4$\times$1800 & 45 & $-$20.00 $\pm$ 0.48\\
2458908.547 & H & 2$\times$1800 & 29 & $-$20.65 $\pm$ 0.80\\
2458910.524 & H & 2$\times$1800 & 28 & $-$18.10 $\pm$ 0.70\\
2458918.493 & H & 4$\times$1800 & 18 & 5.68 $\pm$ 1.50\\
2458921.413 & H & 2$\times$1800 & 13 & 16.72 $\pm$ 1.71\\ \hline
    \end{tabular}
\end{table}

\section{Spectral  analysis}\label{sec:SpTanalysis}

\subsection{The stripped primary and the light ratio}\label{subsec:Bprim}

Without eclipses or further constraints on photometry, the spectra can only be disentangled up to a scaling factor that reflects the light ratio of the two components. 
Before determining the parameters of the two stars, it is necessary to estimate the contribution of each star to the total flux. 
We find that for the primary, the depths of the narrow lines belonging to heavy elements (e.g. S, Si, Fe) are consistent with solar abundance when scaling the primary spectrum to a contribution of about 55\%, implying a 45\% contribution for the Be secondary.

\begin{figure}\centering
    \includegraphics[width=0.5\textwidth]{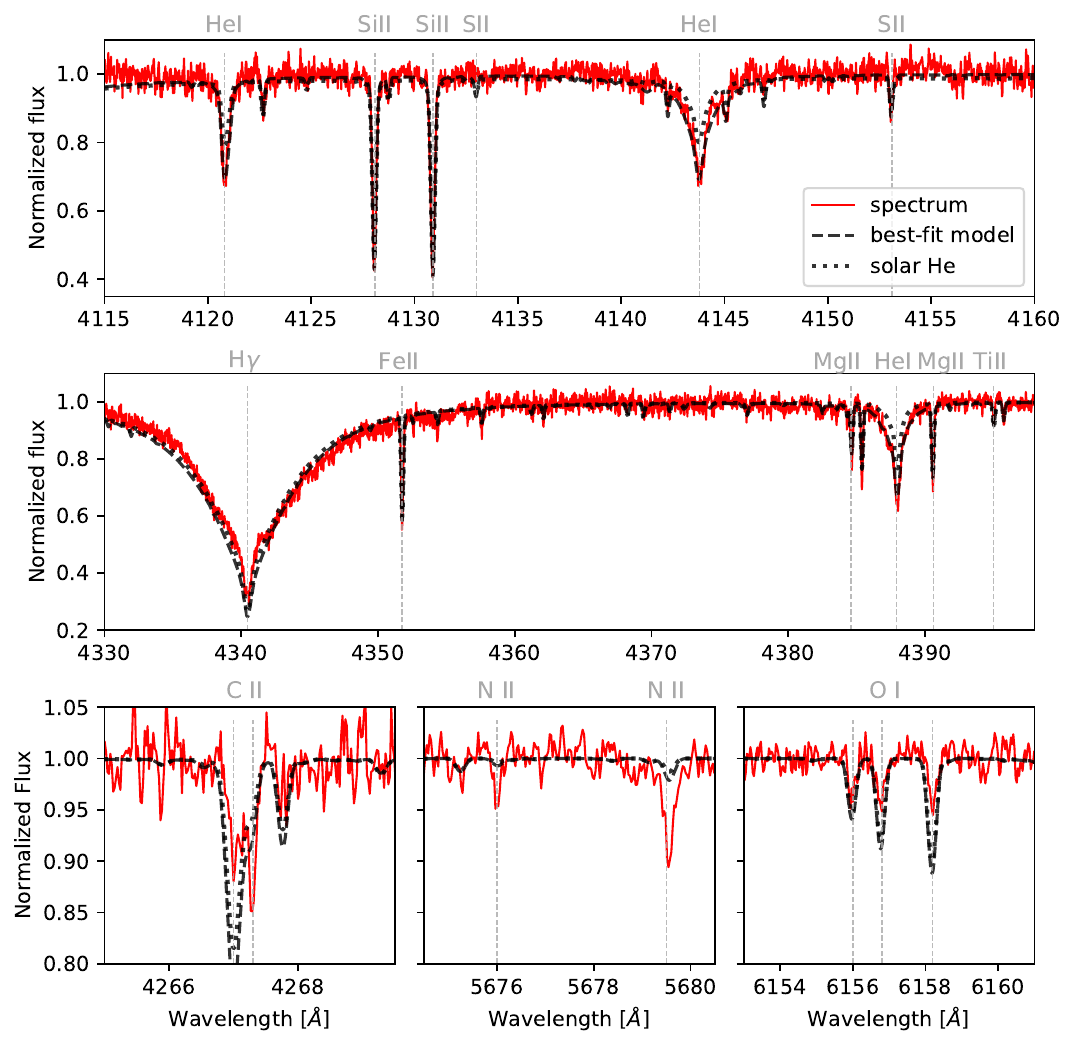}
    \caption{Comparison between the disentangled spectrum of the primary with the best-fit GSSP model and a model of identical parameters but a solar He abundance (see legend). The three bottom panels focus on C, N, and O lines. The main spectral lines are indicated.}
    \label{fig:Bcomp}
\end{figure}

The disentangled, scaled spectrum of the primary is analysed using the Grid Search in Stellar Parameters (GSSP) tool \citep{Tkachenko2015}, constructed with the LTE radiative transfer code \textsc{synthv} \citep{Tsymbal1996} from a grid of \textsc{LLmodel} atmospheres \citep{Shulyak2004}. 
Given the degeneracy in the parameters, we perform the analysis based on the following procedure: we fix the metallicity to solar and set the microturbulence to $\varv_\mathrm{micro}\,=\,2\,$\kms~while we fit for the macroturbulent velocity $\varv_\mathrm{macro}$ and the projected rotational velocity $\varv\,\sin i$.
In a second step, we use diagnostic line ratios such as He\,{\sc i} to Mg\,{\sc ii} and Si\,{\sc ii} to Si\,{\sc iii}  to constrain $T_\mathrm{eff}$, and we rely on the wings of the Balmer and He\,{\sc i} lines to estimate $\log g$. Finally, we vary the He abundance to obtain a best-fit model that reproduces the strength of the He lines. In Fig.\,\ref{fig:Bcomp}, we compare the observed spectrum to the best-fit model as well as a model with the same parameters but a standard He abundance. Our best-fit estimates are given in Table\,\ref{tab:Bstar_properties}.
 Typical GSSP errors are 2~kK on $T_\mathrm{eff}$ and 0.2~dex on $\log g$.

 We confirm the comparably low $T_{\rm eff}$ and $\log g$ values that were reported by \citet{Irrgang2020}. Furthermore, we confirm the more widely accepted low rotational velocity reported previously. Our consistent fit implies that the sub-solar metallicity derived by \citet{Irrgang2020} for heavy elements was a spurious result caused by the dilution of the secondary star. Unlike \citet{Irrgang2020}, we can consistently fit the wings and cores of the He\,{\sc i} lines, having removed the contribution of the secondary component.

Despite the overall solar metallicity, we find that the number density ratio of He to H is about three  times the solar value. Moreover, we confirm strong evidence of CNO-processed material (N enhanced, C and O  depleted) as reported by \citet{Irrgang2020}. This is evident from Fig.\,\ref{fig:Bcomp}, which shows a similar trend in the comparison of the observed CNO lines with the ones predicted by the best-fit model.  The Balmer-line emission (see Fig.\,\ref{fig:ShiftnAdd}) is not considered in this analysis. It may originate in a wind or a disk around the stripped star, but it could also be indicative of a perturbation in the Be-star disk that mimics the primary's orbit \citep[e.g. due to the irradiation of the Be disk by the primary][]{Liu2020}.



\subsection{The Be-type secondary}\label{subsec:Besec}

\begin{figure}\centering
    \includegraphics[width=0.5\textwidth]{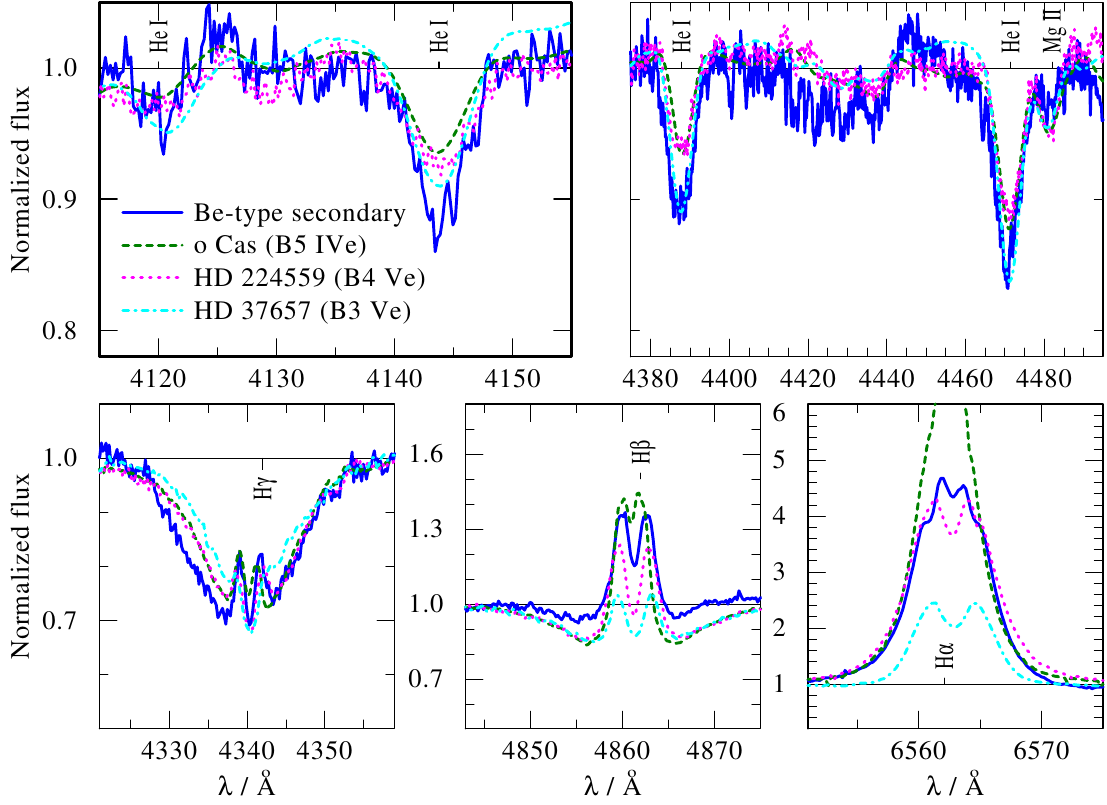}
    \caption{Comparison between the disentangled spectrum of the Be secondary and other Be-type stars (see legend and text). The spectra are binned at $\Delta \lambda = 0.2\AA$ for clarity. }
    \label{fig:Becomp}
\end{figure}

To classify the Be-type secondary, we searched the HERMES archive for representative Be-type stars. Figure\,\ref{fig:Becomp} shows a comparison to HERMES observations of three Be stars: \object{o Cas} \citep[B5\,IVe, ][]{Slettebak1982}, \object{HD\,224559} \citep[B4\,Vne, ][]{Herman1959}, and \object{HD\,37657} \citep[B3\,Ve, ][]{Guetter1968}. Since the spectral lines of \object{o Cas} and \object{HD\,37657} are somewhat narrower than observed for the secondary, we convolve their  absorption-line profiles with rotational profiles that match the observed width of the Be secondary's He\,{\sc i} lines. 
The equivalent widths (EW) of the He\,{\sc i} lines peak around B2-B3\,V, as is also evident from Fig.\,\ref{fig:Becomp}. Since the light ratio was constrained in Sect.\,\ref{subsec:Bprim}, we can use the absolute EW of the He\,{\sc i} lines to obtain a rough estimate of the spectral type. Figure\,\ref{fig:Becomp} indicates that a reasonable match with the He\,{\sc i} and Mg\,{\sc ii} lines is reached for a B3\,Ve star. Hence, we classify the secondary as B3\,Ve. 

The Balmer emission depends primarily on the properties of the disk (e.g. density, inclination). As Fig.\,\ref{fig:Becomp} shows, the morphology of the Balmer lines is best matched with the B4\,Ve star \object{HD\,224559}. The H$\gamma$ line is stronger than typically observed, while H$\beta$ shows little absorption compared to the other stars. However, these slight discrepancies are probably a result of the disentangling and normalisation procedure. 

\begin{figure}\centering
    \includegraphics[angle=90, width=0.5\textwidth]{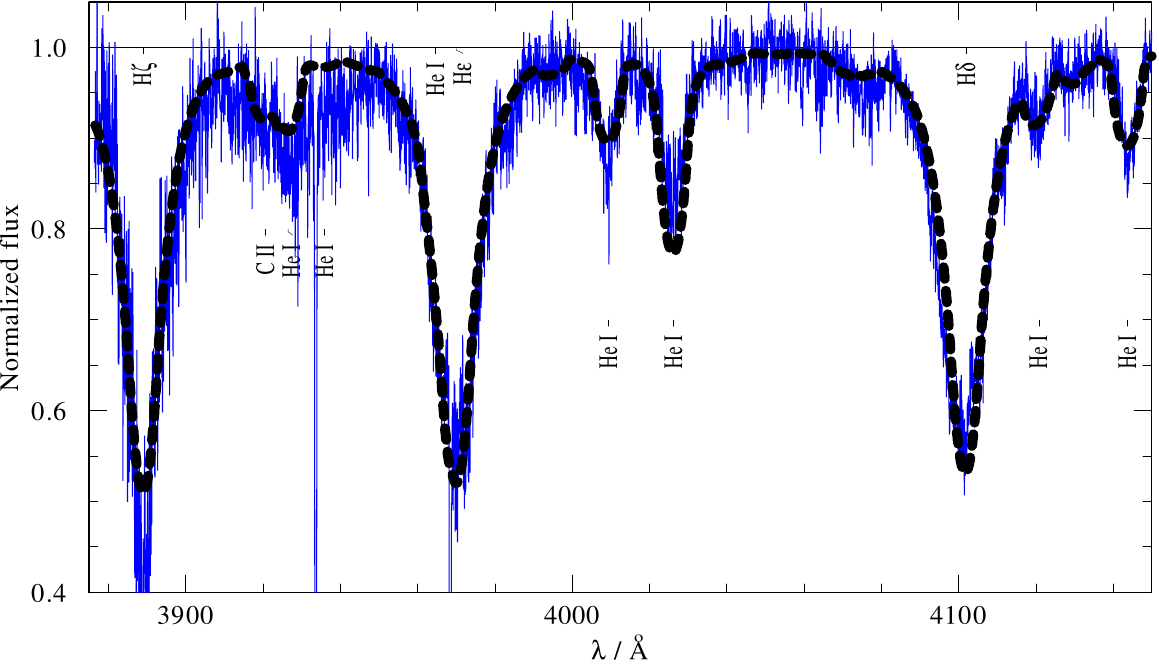}
    \caption{Comparison between the disentangled spectrum of the Be secondary (blue) 
    with a PoWR model calculated with the parameters given in Table\,\ref{tab:Bstar_properties} (black dashed line). The region shown is least affected by disk emission. }
    \label{fig:Beanalysis}
\end{figure}

We refrain from utilising an LTE analysis for the hotter Be secondary, where non-LTE effects may become more important.  Instead, 
we estimate its physical parameters by comparing its absorption spectrum to synthetic spectra calculated with the non-LTE Potsdam Wolf-Rayet (PoWR) atmosphere code \citep{Hamann2003, Sander2015}, relying on pre-calculated grids \citep{Hainich2019} extended for our purpose. We find a good match for \mbox{$T_{\rm eff}{=}18{\pm2}$\,kK}, \mbox{$\log g{=} 4.0{\pm0.3}$\,[cgs]}, and \mbox{$\varv \sin i{=}300{\pm 50}\,$\kms} (Fig.\,\ref{fig:Beanalysis}). 
The reddening and luminosities of the two components are estimated by fitting the observed spectral energy distribution of \Starname~to the sum of two PoWR models calculated with the parameters given in Table\,\ref{tab:Bstar_properties}, assuming the Gaia distance of 2.13\,kpc \citep{Bailer-Jones2018}. 
Despite being fainter in the visual, the Be secondary is the more luminous component due to its higher temperature. We note that the comparable light contribution of the two components provides a natural explanation for the lack of evidence of binary motion in the Gaia astrometry \citep[see][]{Simon-Diaz2020}.


\end{document}